\documentclass[sigconf]{acmart}
\AtBeginDocument{%
  }

\usepackage{wrapfig}
\usepackage{tabularx}
\usepackage{subfigure}
\usepackage{subcaption}
\usepackage{ragged2e}
\usepackage{todonotes}
\usepackage{xcolor}
\usepackage[utf8]{inputenc}         
\usepackage{enumitem}               
\newcommand\im[1]{\textcolor{black}{#1}}
\copyrightyear{2025}
\acmYear{2025}
\setcopyright{cc}
\setcctype{by}
\acmConference[FAccT '25]{The 2025 ACM Conference on Fairness, Accountability, and Transparency}{June 23--26, 2025}{Athens, Greece}
\acmBooktitle{The 2025 ACM Conference on Fairness, Accountability, and Transparency (FAccT '25), June 23--26, 2025, Athens, Greece}\acmDOI{10.1145/3715275.3732218}
\acmISBN{979-8-4007-1482-5/2025/06}
\begin{document}

\title{“I Hadn't Thought About That”: Creators of Human-like AI Weigh in on Ethics \& Neurodivergence} 



\author{Naba Rizvi}
\affiliation{%
  \institution{University of California, San Diego}
  \city{La Jolla}
  \state{California}
  \country{USA}}
\email{nrizvi@ucsd.edu}

\author{Taggert Smith}
\affiliation{%
  \institution{University of California, San Diego}
  \city{La Jolla}
  \state{California}
  \country{USA}}

\author{Tanvi Vidyala}
\affiliation{%
  \institution{University of California, San Diego}
  \city{La Jolla}
  \state{California}
  \country{USA}}

\author{Mya Bolds}
\affiliation{%
  \institution{University of California, San Diego}
  \city{La Jolla}
  \state{California}
  \country{USA}}

\author{Harper Strickland}
\affiliation{%
  \institution{University of California, San Diego}
  \city{La Jolla}
  \state{California}
  \country{USA}}

\author{Andrew Begel}
\affiliation{%
  \institution{Carnegie Mellon University}
  \city{Pittsburgh}
  \state{Pennsylvania}
  \country{USA}}

\author{Rua Williams}
\affiliation{%
  \institution{Purdue University}
  \city{West Lafayette}
  \state{Indiana}
  \country{USA}}

\author{Imani Munyaka}
\affiliation{%
  \institution{University of California, San Diego}
  \city{La Jolla}
  \state{California}
  \country{USA}}
\email{drmunyaka@ucsd.edu}

\renewcommand{\shortauthors}{Rizvi et al.}

\begin{abstract}
Human-like AI agents such as robots and chatbots are becoming increasingly popular, but they present a variety of ethical concerns. The first concern is in how we define humanness, and how our definition impacts communities historically dehumanized by scientific research. Autistic people in particular have been dehumanized by being compared to robots, making it even more important to ensure this marginalization is not reproduced by AI that may promote neuronormative social behaviors. Second, the ubiquitous use of these agents raises concerns surrounding model biases and accessibility. In our work, we investigate the experiences of the people who build and design these technologies to gain insights into their understanding and acceptance of neurodivergence, and the challenges in making their work more accessible to users with diverse needs. Even though neurodivergent individuals are often marginalized for their unique communication styles, nearly all participants overlooked the conclusions their end-users and other AI system makers may draw about communication norms from the implementation and interpretation of humanness applied in participants’ work. This highlights a major gap in their broader ethical considerations, compounded by some participants’ neuronormative assumptions about the behaviors and traits that distinguish “humans” from “bots” and the replication of these assumptions in their work. We examine the impact this may have on autism inclusion in society and provide recommendations for additional systemic changes towards more ethical research directions.\end{abstract}

\ccsdesc[500]{Human-centered computing~Accessibility}
\ccsdesc[500]{Computing methodologies~Artificial intelligence}
\ccsdesc[300]{Social and professional topics~Computing profession}

\keywords{neuronormativity, neuroinclusivity, humanization, humanness in AI, human-like AI}

\maketitle
\section{Introduction}
As human-like AI-powered communication agents such as robots and chatbots become increasingly popular, it is important to examine the ethical concerns surrounding the ways \im{human-like behavior and characteristics or} \textit{humanness} is defined and implemented in such technologies. Our approach to this follows the recommendations of prior research and applies relational ethics in combating algorithmic injustices ~\cite{birhane2019algorithmic}. To this end, we analyze foundational beliefs informing algorithmic design, question what is considered “normal”, and how these definitions may marginalize certain groups~\cite{birhane2019algorithmic}. In particular, we examine how perceptions of humanness may marginalize autistic people, as prior work has shown media representations of autism impact public perceptions, and more positive and accurate representations may even help combat the explicit biases others' hold toward them~\cite{mittmann2024portrayal,jones2021effects}.


The \textbf{double empathy problem} suggests the misunderstandings in communication among autistic and non-autistic people is a two-way issue and not a result of a deficits among autistic people~\cite{milton2012ontological}. In contrast, \textbf{neuronormativity} is the positioning and privileging of neurotypical social behaviors as the “norm”, thus introducing a power imbalance between people with different neurotypes, such as those who are autistic~\cite{legault2024breaking}. Through a neuronormative perspective, autistic people are marginalized for lacking “social skills”. However, in alignment with the double empathy problem, prior work has shown that in neurotype-matched interactions, autistic people have high interactional rapport with other autistic individuals, suggesting that they have their own communication preferences instead of having a deficit of the communication behaviors commonly found in neurotypicals ~\cite{crompton2020neurotype}. Although neuornormative beliefs lead to the dehumanization and marginalization of autistic people, punitive measures, and other epistemic injustices, they continue to be pervasive in our society ~\cite{benson2023perplexing, legault2024breaking, catala2021autism}. For autistic individuals, neuronomativity has often resulted in dehumanizing comparisons to robots, animals, and other non-human entities, which compounds the importance of ensuring technological agents do not reproduce these stereotypes ~\cite{rizvi2024robots, williams2021misfit}. 

Prior work in computing research has examined the ways in which technologies may fail to align with the needs of autistic and other neurodivergent (ND) individuals, or contribute to their marginalization. Autistic people have been notably excluded from the design of robots, which has resulted in robots often replicating dehumanizing stereotypes in their interactions with autistic people ~\cite{rizvi2024robots}. While researchers have identified the unique needs, exclusion, and marginalization of autistic people in other areas of computing research and highlighted the ways in which technology can be made more accessible for them~\cite{van2023understanding, 10.1145/3344919, baillargeon2024puts, begel2020lessons, rizvi2024robots, taylor2023co, hijab2024co, zolyomi2021social, 10.1145/3569891, guberman2023not}, there is a notable gap in research exploring the perspectives of the people who create such technologies, particularly their intentionality, perceived and intended impact, and other ethical considerations of their work. Additionally, it remains unclear how the creators of these AI systems conceptualize and operationalize humanness in their work, and whether these efforts perpetuate, \im{support}, or challenge neuronormativity.

To address this gap, our work examines the beliefs and experiences of researchers, designers, and engineers who work on human-like AI technologies (i.e. robots and chatbots). We conduct a qualitative study using an interpretative phenomenological analysis of two interviews and two surveys from each individual participant in our study. Our work also examines the alignment of their work with neuronormative standards of communication, and the barriers they face in making their work more accessible and inclusive.

To this end, we investigate the following:

\begin{itemize}
    \item How do AI system makers conceptualize and implement “humanness” in their systems?
    \item \im{From the developer perspective}, what are the \im{potential} ethical and societal impacts of designing AI systems to mimic human communication? 
    \item How do AI makers' considerations toward replicating human behavior reinforce neuronormative standards and marginalize autistic individuals?
    \item What challenges prevent AI developers from incorporating neurodiversity and accessibility into their design processes, and how can these be addressed?
\end{itemize}

\section{Related Work}
Advancements in AI increasingly aim to replicate human qualities, yet ethical concerns arise from how “humanness” is defined, often overlooking the experiences of historically dehumanized groups such as autistic individuals. In this section, we explore how humanization is implemented in human-like AI agents such as robots and chatbots, the anti-autistic biases prevalent in AI, and the ways in which such technologies may perpetuate stereotypes or fail to align with the needs of autistic people.
\subsection{Humanizing AI}
\subsubsection{Origins} Alan Turing proposed a simple and observable test to evaluate whether a machine can exhibit human-like intelligence through conversation~\cite{turing2009computing}. The Turing Test, which ssesses whether or not a judge can reliably distinguish between answers from a human and a machine~\cite{turing2009computing}, is considered a benchmark for assessing AI’s humanness. However, its emphasis on immitating human behavior raises ethical concerns about which set of humans are being treated as the gold standard for 'humanness', especially as the test considers it to be a key benchmark for intelligence ~\cite{murugesan2025turing, jones-bergen-2024-turing}. This is particularly relevant for marginalized groups, such as autistic individuals, who may be unfairly measured against neuronormative social expectations that AI is also trained to replicate. Despite these concerns, the test has continued to influence decades of AI research on natural language and cognition, from early chatbots to modern dialog models ~\cite{shum2018eliza,xue2024bias,jones-bergen-2024-turing,murugesan2025turing}.

\subsubsection{Contemporary Work} From the Turing Test \cite{turing1950} through Weizenbaum’s ELIZA \cite{weizenbaum1966} and onward to Breazeal’s sociable robots \cite{breazeal2003}, the drive to humanize technology has spanned domains from natural language processing to embodied robotics. It has shaped contemporary research on how robotic appearances and gestures evoke emotional responses, and continues to inform contemporary HCI work aimed at cultivating trust through empathetic yet transparent AI \cite{cuadra2024, hauptman2022}. Designers now consider anthropomorphic (i.e. human-like) behaviors such as voice, facial expressions, or gestures with ethical frameworks emphasizing user autonomy, cultural sensitivity, and disclosure of AI’s limitations \cite{churchill2024, fenwick2022}. While researchers are increasingly moving toward responsible AI research, such work continues to overlook people with disabilities, particularly those who are neurodivergent.

\subsubsection{Limitations} Recent works continue to build upon Turing's foundational ideas. For example, one study on conversational agents highlighted that incorporating cognitive, relational, and emotional competencies can enhance user engagement and lead to more human-like capabilities in these agents ~\cite{chandra2022or}. Similarly, another paper highlighted the modalities (verbal, non-verbal, appearance) and footing (similarity and responsiveness) that can help optimize interactions with end-users by moving toward humanness ~\cite{van2020human}. However, a systematic literature review highlighted many ethical concerns with humanizing AI, such as the usage of AI to manipulate end-users, for example, by influencing their voting decisions ~\cite{abraham2023systematic}. Importantly, this work does not address the ethical implications of how humanness is defined and implemented in AI. This issue is particularly significant for autistic people who have been historically dehumanized due to their cognitive, relational, or emotional differences ~\cite{pearson2021conceptual} and have been unfairly compared to robots, with robots frequently and paradoxically being used as mentors to guide them toward conforming to a constructed notion of “humanness” ~\cite{williams2021misfit, rizvi2024robots}.

\subsection{Accessibility Considerations}
A systematic review of chatbot accessibility organized considerations into five categories: content, user interface, integration with other web content, developer process and training, and testing ~\cite{stanley2022chatbot}. This work highlighted the importance of ensuring chatbots use straightforward and literal language with a single-minded focus for broader accessibility~\cite{stanley2022chatbot}. Similarly, another review detailed accessibility concerns in chatbots, and how they arise from a lack of support for alternative input methods and assistive technologies, lack of clarity and consistency, and a lack of simplicity ~\cite{10306062}. The recommendations provided in this study included providing navigational assistance, keyboard shortcuts, and feedback while reducing disruptive factors to improve accessibility ~\cite{10306062}. A similar study on pre-existing accessibility guidelines for human-robot interaction research found that some researchers had considered ad hoc guidelines in their design practice, but none of them showed awareness of or applied the guidelines in their design practices ~\cite{qbilat2021proposal}. Notably, these guidelines have a lot of overlap with the ones proposed for chatbots, for example, both encourage using multiple modals, providing support for assistive technologies, navigation assistance, and feedback ~\cite{10306062, qbilat2021proposal}. 


\subsection{Anti-Autistic Biases in AI}
Prior studies have examined several ways in which AI displays biases through applying neuronormative standards in their classification of emotions and behaviors, and stereotypical representations of autism. Notably, one study found training AI on neurotypical data leads to a mismatch with the needs of autistic users, highlighting the limitations of datasets that do not include accurate representations of autistic people~\cite{begel2020lessons}. Similarly, prior work has examined the ways in which AI-powered emotion recognition in speech may perpetuate ableist biases through their focus on pathologizing the communication behaviors of autistic people, especially as they build upon foundational work dehumanizing autistic people by comparing them to computers~\cite{10.1145/3593013.3594011}. Additionally, diagnostic voice analysis AI may misclassify autistic people’s speech as “atypical” or “monotonous” which results in the system making ableist assumptions ~\cite{10.1145/3593013.3594032}. AI-powered talent acquisition systems may also misunderstand autistic people's behaviors by judging them based on neurotypical standards, resulting in misclassifying their lack of eye contact as a lack of confidence, for example, and negatively impacting their overall perceptions of autistic candidates ~\cite{10.1145/3531146.3533169}. Even generative AI has a tendency to produce biased representations of autistic people through stereotypical emotions such as anger and sadness and engagement in solitary activities ~\cite{wodzinski2024visual}.

\subsection{Robots, Chatbots, and Autism}
Anti-autistic biases are prevalent even in chatbots, as one study uncovered that GPT-4 displays biases toward resumes that mention disabilities, including autism~\cite{10.1145/3630106.3658933}. Additionally, a systematic review of human-robot interaction research found robots may marginalize autistic people through a power imbalance in their user interactions by acting as mentors who help autistic people move toward humanness~\cite{rizvi2024robots}. Interestingly, there is a notable contrast between the preferences of autistic people and others when it comes to robots and chatbots---prior studies found that autistic people have a preference for non-anthropomorphic robots and chatbots ~\cite{ko2024chatbot, rizvi2024robots, ricks2010trends}, while anthropomorphized chatbots and robots may increase satisfaction for other users ~\cite{klein2023impact}, particularly those who have a higher desire for human interaction ~\cite{sheehan2020customer}. This suggests that even the decision to implement human-like traits in a bot may marginalize autistic users by overlooking their preference in favor of positive user experiences for other groups.

\subsection{Creator Perspectives on Ethics and Limitations}
Prior work exploring the perspectives of creators suggests that ethical concerns are not purposefully overlooked, but are rather the result of other practical limitations that can be addressed to improve the alignment of theoretical ethical beliefs with real-world practices. For example, one study investigated the ethical caveats of conversational user interfaces by interviewing both the designers and end-users of chatbots, and identified mismatches between the designers' user-centered values, and the resulting user experiences shaped by technical constraints in the real-world ~\cite{Mildner_2024}. Similarly, another study examined the thoughts of design leaders toward implementing ethics in commercial technology settings found that although the designers recognize the importance of inclusive design, they face limitations due to the pressure to deliver products quickly, which creates a gap between their aspirational ethical guidelines and their real-world projects ~\cite{Lindberg_2023}. Other works identifying the needs of machine learning practitioners, designers, and data practitioners has highlighted the importance of practical tools and guidance in helping bridge the gap between literature on fair ML research and real-world constraints ~\cite{Holstein_2019}, as unclear best practices makes them struggle with ``ethical correctness'' ~\cite{Dhawka_2025}.

\section{Methods}
Our study consisted of a comprehensive methodological approach encompassing virtual semi-structured interviews, strategic participant recruitment, and rigorous data analysis to explore participants’ experiences with creating human-like AI, and their perspectives on intersectionality, accessibility, and neurodiversity accommodations.

\subsection{IRB Approval and Other Ethical Concerns}
Our IRB was approved by the appropriate Institutional Review Board (IRB). Prior to participating in each portion of our study, the participants filled out a consent form that provided details on the time commitment, participant and researcher expectations, the nature of the data being collected, the compensation being provided, and other information about the research team. Participants were explicitly asked for permission to record their interviews, and were made aware that they could quit the study at any time at their own discretion. 

\subsection{Participant Recruitment}
Our recruitment strategy consisted of purposive sampling, snowball sampling, and outreach through social media. We recruited participants based on specific eligibility criteria, including being U.S.-based, at least 18 years old, and employed as researchers, designers, or engineers working in the development of human-like AI systems. We used purposive sampling to target a pool of 154 professionals from U.S.-based tech companies and universities specializing in AI and user experience (UX). These professionals were found through a manual search on the websites of their employers. We also reached out to community-based organizations to recruit participants, such as affinity groups supporting underrepresented communities in computing and AI, nonprofits advocating for disabled and queer individuals in technology fields, and employee resource groups at tech companies. Finally, we used snowball sampling by inviting eligible researchers from our networks and those recommended by other participants. We posted advertisements on social media platforms, including LinkedIn and Twitter. Ultimately, 16 individuals qualified for and completed the full study. 

We define ``human-like AI” as systems mimicking human communicative or behavioral cues (e.g., text, speech, gestures). We include participants working on self-described human-like technologies (e.g. robots, chatbots, digital avatars). We did not use the phrase ‘human-like’ during our recruitment to avoid recruitment bias. Here is an example of a passage used in our recruitment:

\begin{quote}
        ``We are interested in learning about the background and experiences of professionals in AI, Machine Learning, and/or Data Science. We are specifically interested in learning more about the processes used during the development of human-facing tools and the life experiences of those professionals.''

\end{quote}

\subsection{Study Design}
We conducted semi-structured interviews virtually, with each session lasting between 30 and 45 minutes. The study was divided into two sections, with each section having a survey and an interview. The first section collected data on the participants’ demographics and dived deeper into their experiences with working on a human-like AI project. The second section focused more on their views of diversity, disability, and neurodiversity. Due to the nature of the questions presented in the second part, we alternated the order in which the survey and interview were presented to each participant to provide counterbalance. Participants were compensated \$60 for their participation, provided in two installments: \$30 after completing each section.

The interview and survey questions were carefully-designed to minimize response bias and gain a deeper understanding of participant perspectives. A collaborative team of researchers with expertise in software engineering, responsible AI, and security/privacy research developed the study materials. The interview and survey questions are available in the Appendix ~\ref{Appx}.

\subsubsection{Surveys}
The study included two surveys, which can be found in our Appendix sections ~\ref{Survey1} and \ref{Survey2}. The first survey gathered demographic information, including age, gender, socioeconomic status, education level, disability, race, and marital status with questions taken from previous studies ~\cite{Windsor2015,Ruberg2020,Scheim2019,Spiel2019,USDHHS2018,Fallah2017,Lee2020}. The second survey explored participants’ views on topics such as intersectionality, diversity, and workplace accommodations for neurodiversity. 

\subsubsection{Interviews}
The first interview consisted of carefully designed questions on the design, development, and testing of a recent project completed by our participants focusing on human-like AI. This interview also included questions on general accessibility considerations, accessibility for neurodivergent individuals, and the broader impact their project may have on shaping the perceptions of communication norms and the humanness of both their end-users and other AI creators who build on their work. The questions for this section of the interview are available in our Appendix Section ~\ref{Int1}.

The second interview focused on understanding each participant’s views of intersectionality, diversity, disability, and neurodiversity. In this interview, we also presented each participant with scenarios of workplace interactions between colleagues of diverse backgrounds including different neurotypes that were adapted from a previous study ~\cite{rizvi2021inclusive}. The identities of the people in the scenarios were not revealed to the participants to avoid response bias, but were indirectly communicated through their dialogue focusing on sensory and communicational differences common in individuals with ADHD and autism. These questions were designed to dive deeper into the participants’ knowledge and acceptance of neurodiverse identities. The questions for this section of the interview are in our Appendix Section ~\ref{Int2}.


\subsection{Analysis}
We used an interpretative phenomenological analysis (IPA) approach to examine the data collected \cite{eatough2017interpretative}. IPA is a qualitative method that explores how individuals perceive and interpret their lived experiences, typically involving iterative steps such as immersing oneself in the data, generating initial codes, searching for emergent themes, and synthesizing these into interpretative narratives ~\cite{eatough2017interpretative}. The method also requires researchers to `bracket' or set aside their biases at the beginning of the analysis, and have a data validation process to ensure the analysis results reflect the participants' experiences and perspectives ~\cite{smith2004reflecting}. We used this method to gain a better understanding of each participant’s experiences as they were from diverse professional backgrounds and were working on different kinds of projects. Additionally, this analysis method benefits from quality and depth over quantity and breadth, which allows us to concentrate more deeply on the data collected in our multiple interviews and surveys from each individual ~\cite{eatough2017interpretative}. 

The analysis was conducted by a team of four researchers (authors 1-4), with the findings validated by two additional researchers (authors 5 and 6) specializing in neurodiversity and accessibility. All interviews were recorded and transcribed with the participants’ knowledge and consent. To ensure an unbiased approach, each researcher engaged in a bracketing process by documenting their own perspectives and biases before beginning the analysis. This process was used to help mitigate potential personal biases on the findings. Each researcher independently reviewed the interview transcripts, taking verbatim notes with direct quotes to capture the participants’ experiences. The team then engaged in iterative group discussions to summarize findings, identify recurring patterns, and generate codes. These codes and themes were further refined collaboratively and validated by external experts to ensure the accuracy and rigor of the analysis.

\subsection{Positionality}
This work was led by an autistic researcher with a neurodiverse research team. Although the researchers have diverse racial, ethic, and gender backgrounds, we are all English-speaking and US-based. Thus, our work has limitations due to our Western and Anglo-centric perspective. We acknowledge that due to these limitations, our findings may not be applicable to cultures and languages, and encourage future work to explore other perspectives.
\section{Results}
\begin{wrapfigure}{r}{0.5\linewidth}
    \vspace{-10pt} 
    \centering
    \includegraphics[width=\linewidth]{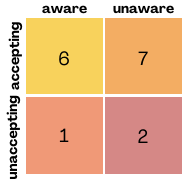}
    \caption{An overview of our participants' knowledge and acceptance of neurodivergence.}
    \label{fig:NDKnowledge}
    \vspace{-6pt}  
\end{wrapfigure}

We present insights from our interviews and surveys on the experiences and beliefs of researchers, designers, and engineers working on human-like AI. In particular, we focus on their views toward the accessibility and ethical implications of their work, and its alignment with neurodiversity. Examples of their projects include robots, chatbots, video games, and assistive technologies. Table ~\ref{tab:participant-demographics} provides an overview of our participants’ demographics while Figure ~\ref{fig:NDKnowledge} summarizes their knowledge and acceptance of the behaviors and traits common in neurodivergent individuals that may be considered atypical by neuronormative communication standards ~\cite{rizvi2021inclusive, wise2023we}. The participants were classified as unaccepting, accepting, unaware, or aware based on their responses to our scenario-based questions in the second interview. For example, if a participant answered, “that’s weird”, when asked to imagine a colleague who wears headphones, but said “that’s ok” when we specified the individual does it to avoid sensory overload, they were classified as being unaccepting but aware of neurodivergence as they initially displayed bias toward the behavior but understood it may just be a behavioral difference.


\begin{table*}[!ht]
\centering
\begin{tabular}{|c|c|c|c|c|c|c|c|}
\hline
\textbf{ID} & \textbf{Gender} & \textbf{Age} & \textbf{Race} & \textbf{Education} & \textbf{Disability} & \textbf{ND} & \textbf{Areas} \\ \hline
1 & M & 31-40 & White & PhD & None & Yes & AI \\ \hline
2 & NB & 18-22 & 2 or more races & Bachelors & None & Yes & AI \\ \hline
3 & F & 31-40 & Other & No degree & Has disability (unspecified) & No & HRI \\ \hline
4 & M & 23-30 & Asian & Bachelors & None & No & AI \\ \hline
5 & M & 51-60 & Black & PhD & Blind & No & HCI \\ \hline
6 & F & 23-30 & Asian & PhD & None & No & HCI \\ \hline
7 & F & 23-30 & Asian & Bachelors & None & No & AI \\ \hline
8 & M & 23-30 & Asian & PhD & Deaf & No & HCI \\ \hline
9 & F & 23-30 & Other & PhD & None & No & HCI \\ \hline
10 & M & 31-40 & White & PhD & None & No & HCI \\ \hline
11 & M & 23-30 & Hispanic/Latino & PhD & None & No & AI \\ \hline
12 & F & 23-30 & Asian & Bachelors & None & No & HCI \\ \hline
13 & M & 23-30 & 2 or more races & PhD & None & No & AI \\ \hline
14 & M & 51-60 & White & PhD & None & No & HCI \\ \hline
15 & M & 31-40 & Asian & PhD & None & No & AI \\ \hline
16 & M & 31-40 & Hispanic/Latino & PhD & None & No & HRI \\ \hline
\end{tabular}
\caption{The demographics of the participants in our study. ND refers to neurodivergence.}
\label{tab:participant-demographics}
\end{table*}


\subsection{Desirable Traits}
Through a qualitative thematic analysis of the participants’ responses during our interviews, we uncovered 12 desirable traits they implement to make their technologies appear more human-like. Table ~\ref{tab:personasdetailed} details the prevalence of each trait, with “personalized”, and “uses natural language” being the most popular desirable traits. Notably, traits such as “ethical”, and “compassionate” were less common.

\begin{table*}[htp] 
\centering

\label{tab:personasdetailed}
\begin{tabularx}{\linewidth}{|l|c|>{\RaggedRight\arraybackslash}X|} 
\hline
\textbf{Desirable traits} & \textbf{Prevalence} & \textbf{Examples} \\ \hline
Personalized & 10 & Mimics interactions with family members, peers, or friends \\ \hline
Uses natural language & 7 & Delayed responses, multimodal communication, organic conversations \\ \hline
Serious & 6 & Professional tone, not overly positive, not funny \\ \hline
Simple & 6 & Does not contain unnecessary features \\ \hline
Helpful & 5 & Offers clarifications, task-oriented, time-saving tool \\ \hline
Encouraging & 3 & Perky, has a rewards system \\ \hline
Compassionate & 3 & Friendly, empathetic, sympathetic \\ \hline
Therapeutic & 3 & Mimics warm nurse and interactions with therapists, calming \\ \hline
Emotive & 2 & Expresses emotions through vocal and facial cues \\ \hline
Teaches & 2 & Provides casual learning experience, teaches sensitive topics \\ \hline
Ethical & 2 & Safe, private, does not make users feel watched \\ \hline
\end{tabularx}
\caption{An overview of the different desirable traits and their prevalence as implemented by our participants in their bots to make them appear more human.}
\end{table*}

\subsection{Undesirable Traits}
\begin{figure}[htp]
    \centering
    \begin{minipage}[t]{0.32\textwidth}
        \centering
        \includegraphics[width=\linewidth]{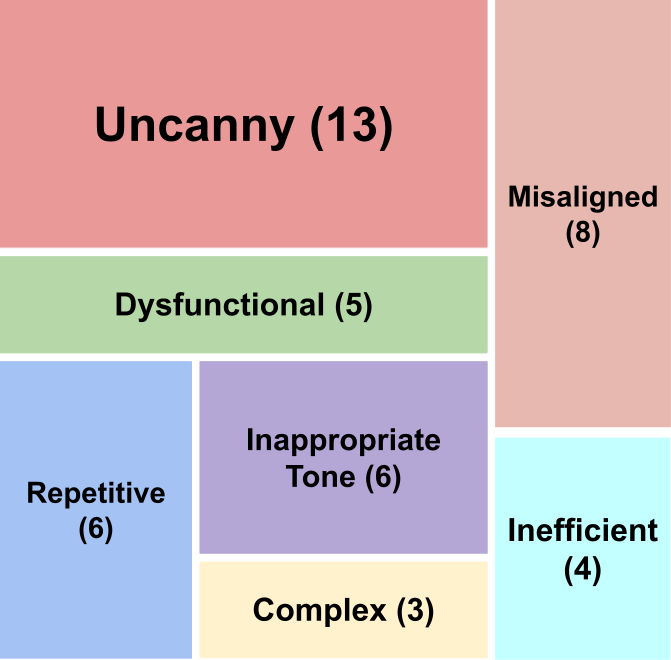}
        \caption{A list of undesirable traits discussed by our participants that may upset or frustrate their users.}
        \label{fig:undesirabletraits}
    \end{minipage}
    \hfill
    \begin{minipage}[t]{0.32\textwidth}
        \centering
        \includegraphics[width=\linewidth]{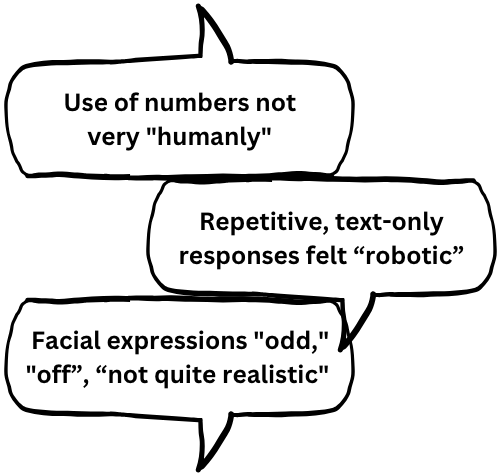}
        \caption{The words used by our participants to refer to undesirable traits and behaviors that frustrated their users.}
        \label{fig:quotes}
    \end{minipage}
    \hfill
    \begin{minipage}[t]{0.32\textwidth}
        \centering
        \includegraphics[width=\linewidth]{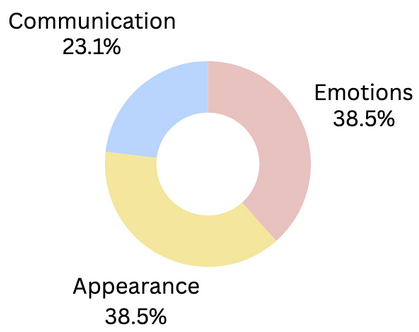}
        \caption{A breakdown of the different aspects of a bot considered uncanny by our participants.}
        \label{fig:uncanny}
    \end{minipage}
\end{figure}

The participants also identified undesirable traits that caused communication breakdowns with their end-users and resulted in the users getting frustrated or upset with the system. These traits, shown in Figure ~\ref{fig:undesirabletraits}, focused on the behaviors, utility, or appearance of the system. A bot would be considered “inefficient”, for example, if it took too long to respond, understand the user, or complete the task specified. Similarly, “dysfunctional” bots were those that offered incorrect answers or experienced other technical difficulties. The participants also expected their bots to have appropriate tone based on the context of the communication, the identity of the user they were interacting with, and the system’s assumed role in the interaction (e.g. as a boss, teacher, or peer). For example, the bots must not appear to be too `child-like', which can be avoided by making them speak `eloquently'. Another source of user frustration reported by our participants was a misalignment of their technologies with the participants’ needs. These misalignments could be cultural, for example, if the bot struggles with a user’s name, or simply due to their irrelevance to the user’s needs. The most commonly reported undesirable trait was the bot appearing “uncanny”. As shown in Figure ~\ref{fig:uncanny}, this was often the result of the system’s communicative behaviors, emotional expression, or appearance. Figure ~\ref{fig:quotes} provides examples of the ways these behaviors and traits were described by our participants.

\subsection{Accessibility} 
Accessibility considerations were mainly discussed by the participants for physical disabilities or globalization (n=12). Participants highlighted challenges such as language barriers and technological limitations, including slower devices. P3 emphasized, “It needs to be accessible to, you know, employees in Malaysia, Denmark, Europe, South America,” while P6 noted issues such as users not owning personal devices and bots struggling to recognize Indian names, leading to user frustration. P1 acknowledged the need for accommodations in their app for medical professionals after drawing a personal connection to a sibling’s vision impairment, noting that “he uses lots of accessibility tools for his study.” However, this recognition only extended to physical disabilities. When asked about accommodations for neurodiverse individuals, the participant stated that they “don't have an answer for this.” Similar conclusions were drawn by several other participants who considered and implemented accessibility for physical disabilities but not neurodiverse conditions (P4, P8, P11). Participant 11, who had implemented colorblind-friendly accommodations for their bot, mentioned that they were “not sure at all” and that “it’s not something I consider” when asked about accessibility for neurodiverse individuals. P4 considered accessibility for blind and low-vision individuals, implementing a screen reader for their application, but stated that they were “not very sure how it would change their interaction with the tool” when it came to neurodiverse individuals. 

\subsection{Neurodivergence}
Our findings reveal that while some participants demonstrated an awareness of the unique traits and differences of neurodivergent individuals (n=7), many had not considered their specific accessibility needs (n=12). For instance, P9 admitted, “I guess those populations are not really my expertise,” while P13 acknowledged, “I actually don't know enough, I guess, about autism to really know how this would affect them.” When prompted, this lack of understanding often led participants to default to design choices that either eliminated features rather than adapting them or made assumptions about neurodivergent users’ preferences. For example, P10 noted that humor could be misinterpreted by neurodivergent users, potentially leading to a poor user experience, and chose not to include it as a feature. Other participants (P7, P8, P16) acknowledged the value of customizing system features for various user groups, but also made assumptions about neurodivergent users’ interaction needs. P16 stated, “[neurodivergent people] need some kind of motivational behavior from the robot in order to encourage them to engage during the interaction,” and P7 suggested “some extra persuasive intervention tools” for neurodivergent people who “suffer from a lack of motivation.” Similarly, P3 assumed their current system would be “super, super friendly” for neurodivergent users, “even if they have like issues with spelling, because we have so many ESL workers, there are some tolerances around [...] not exact matches,” and because the bot “responds with simple things.” None of these participants directly examined the needs of neurodivergent individuals in their studies. In contrast, P14 recognized the significant appreciation from neurodivergent communities for simply being considered during the design process, stating, “For the neurodiverse population [...] the fact that we were designing for them at all [...] there was an outsized appreciation for that. And I think that's basically because they're a commonly neglected or ignored community.”

\subsection{Communication Biases}
Participants also made assumptions about communication preferences and behaviors. Some of these assumptions involved the mode of communication. For example, P11 viewed text-based communication as “overly robotic”. P12 described switching from a numerical scale to using words in the bots’ interactions with the user to “humanize” the experience as numbers felt “unnatural”. Others focused on the tone or content of the communication. For example, P13 perceived overly positive responses as “unnatural”, especially when the bot is prompted to be angry or hateful toward certain communities. Additionally, some participants made assumptions surrounding the impact of certain behaviors. P11 believed making their chatbot repetitive and not providing a direct reward during interactions made it “boring” and “robotic” for the end-users and negatively impacted their engagement. Participants prioritized emotional responsiveness in context of their bots’ conversation and often equated this with humanness. P2's bot was designed to be remain “empathetic” and “grounded” when discussing serious topics to maintain a calming environment for the user. P15 was particularly concerned with how empathy was expressed in the bot’s conversational style and tone, believing “robotic, monotonic advice would make the bot appear less human.

\subsection{Neurotypical Bias in “Friendly” System Design}
When participants discussed their design choices around making systems “friendly” or “personable,” they often defaulted to neurotypical social conventions and communication patterns ~\cite{wise2023we}. For example, P3’s emphasis on making their system “cute, perky, and friendly” reflects assumptions about universal preference for social interaction styles that may not align with neurodivergent users’ needs who prefer more straightforward systems~\cite{robins2006does, rizvi2024robots}. Additionally, some participants’ systems (P3, P5) prominently featured an anthropomorphized face, a design choice likely to be less relevant to some autistic users for whom such conventional facial signifiers of emotion fail to resonate~\cite{zolyomi2021social, robins2006does}. P7 considered incorporating emojis and believed that emotional content was key to humanness, saying of their current system: “these kinds of agents don’t have emotions, so it would probably not be that fluid or, like, more human.” Our participants’ emphasis on emotional facial expression may cause users and researchers to associate a specific mode of communication via facial expressions with humanity, which misaligns with the findings of prior work showing autistic individuals prefer interacting with plain and featureless bots over humanized ones ~\cite{robins2006does}.

\subsection{Ethics}
The ethical concerns discussed by our participants were centered around privacy, model biases, and their perceived responsibilities in creating more ethical and accessible technologies. While participants were questioned about the broader impact of their work, the majority did not discuss any broader ethical considerations.

\subsubsection{Privacy}
Participants expressed varying perspectives surrounding privacy. Some participants were actively working on improving their systems' privacy. For example, P6 described disabling certain features to avoid invading users’ privacy, while P12 highlighted efforts to avoid making users feel surveilled. However, other participants were less concerned with privacy. P10 did not prioritize privacy due to their focus on Gen Z users, observing generational differences in privacy attitudes, “Gen Z [are] less concerned about privacy than previous generations.” 

\subsubsection{Model Bias and Accessibility}
Notably, ethical concerns related to AI model biases were raised by HCI-focused participants (P2, P6, P12) but were absent among AI-focused participants (P1, P4, P13), sometimes explicitly, as one participant remarked `we are machine learning people, not HCI people'. When discussing accessibility, AI-focused participants often downplayed its necessity. For instance, P1 stated, “I don’t know if [the results of] this model are necessary to be accessible because it's more helpful to the doctors and biologists,” assuming that medical professionals do not have accessibility needs. Similarly, P11 remarked, “We weren’t UI people; we were ML people,” highlighting a lack of cross-disciplinary consideration in their designs. Such sentiments were also shared by other participants who believed accessibility is the responsibility of the product team (P4) or will inevitably occur when needed as a result of commercialization (P1). Overall, accessibility and ethics were noted as a future concern or beyond the scope of their responsibilities.

\subsubsection{Broader Ethical Concerns}
A significant concern was the lack of attention to broader ethical implications, particularly in relation to societal norms surrounding communication. Most participants (n=12) did not discuss the downstream effects of their work on either future researchers or end-users, particularly surrounding communication norms. This was especially troubling given the tendency of participants to mimic existing technologies like ChatGPT, as illustrated by P7: “[We] kept it really simple and tried to mimic other chat bots, for example, the ChatGPT interface.” Some of the participants more explicitly mentioned being inspired by communication behaviors that “humanized” other technologies, such as Duolingo’s bird icon which displays facial expressions corresponding to users’ engagement (P12). Yet, while the participants noted being influenced by the communicative behaviors of existing technologies, they did not discuss how their own work may similarly influence others.

\subsection{Barriers}
While the accessibility considerations mentioned by participants, such as having a simple interface, and providing navigational assistance were in alignment with prior work on accessibility standards for robots and chatbots ~\cite{10306062, qbilat2021proposal}, the implementation of these features faced significant barriers, the most common among them being their organization’s priorities (P1, P3, P4, P7, P13, P15).  As P3 explained, “So accessibility is probably not [our company’s] strong suit because it's not a consumer-facing org,” overlooking the potential accessibility needs of employees and other end-users. Other barriers included funding constraints (P12), time limitations (P14), and technical challenges (P10, P11, P16). 

\section{Discussion}

Our findings suggest that human-like AI continues to promote neuronormative standards of communication. We investigate the impact this may have on dehumanizing autistic people and recommend systemic changes to move toward more ethical research.

\subsection{When “Uncanny” Meets Stigma: Parallels with Autistic Stereotypes}
Our participants notably described their perceptions of uncanny qualities in their bots through stereotypes that are often also applied to autistic individuals. In particular, P11 described text-only communication as “too robotic,” due to a lack of human “warmth” or “natural flow” in the system’s interactions. Similarly, autistic people are also perceived as being “robotic” due to differences in their communication behaviors, which include a preference for text-based communication for clarity and reduced social pressure \cite{howard2021anything, williams2021misfit, rizvi2024robots}. This highlights how our participants’ perceptions of “robot-like” communication may inadvertently reinforce problematic assumptions about neurodivergent behavior. Similarly, P15 pointed out that certain body movements and microexpressions seemed “off” or “not quite realistic,” contributing to an overall sense of the system being unnatural or inauthentic. These comments echo broader societal tendencies to dehumanize individuals who express emotions in ways deemed “atypical” \cite{o2020diagnostic}, a common way autistic people are marginalized in our society and are often pressured to mask their natural behaviors despite the negative impact on their well-being \cite{radulski2022conceptualising}.

Further highlighting the ways in which user interactions can be shaped by social stereotypes that correlate with disability stigma, P16 perceived the robot as more of a child than a caregiver. The disconnect between the system’s expected competence and its perceived immaturity caused some user frustration. This echoes the infantilization that autistic people frequently face \cite{stevenson2011infantilizing}, highlighting yet another way in which marginalization ties autistic identities to robot-like attributes \cite{williams2021misfit}. Notably, P15 also drew attention to a broader distinction between humans and robots: the ability to infer another person’s internal state from subtle external cues. Similarly, autistic individuals are often accused of lacking a “theory of mind”— the assumedly universal ability to empathize by putting oneself in another’s position—according to dominant neuronormative standards \cite{smukler2005unauthorized}. These parallel perceptions of “deficient” empathy in both robots and autistic people reinforce the belief that atypical communication or emotional expression is inherently less human. 

These participant insights not only emphasize how easily technology can be perceived as uncanny, but also how such perceptions are linked to normative expectations for communication, emotional expression, and social intuition. In doing so, they also highlight the risk that implementing such expectations in bot design in attempts to avoid the uncanny may reinforce harmful stereotypes about autistic people, who are frequently subjected to similar judgments and dehumanizing labels.

\subsection{Humanizing Machines, Dehumanizing Humans}

Dehumanization, whether subtle or overt, appears with alarming frequency in inter-group relations ~\cite{kteily2017backlash}. Subtler forms involve ascribing fewer human emotions or “complex” traits (e.g. maturity, civility or refinement) to outgroups, while more blatant instances compare marginalized groups to animals, machines, or primitive beings \cite{kteily2017backlash, rizvi2024robots}. Interestingly, in our study, the participants described frustrations with their bots being perceived as ``childlike” or not communicating ``eloquently,” exemplifying the ``complex traits” associated with machines that are also associated with dehumanized human groups. Additionally, our participants often associated behaviors preferred by autistic people, such as text-based communication, with robots, highlighting an explicit dehumanization of neurodivergence \cite{howard2021anything}. 

Such dehumanizing stereotypes can have a serious negative impact on the communities targeted by them, and thus it is important to address them. They may foster hostility, discrimination, and violence, both systemic and overt \cite{kteily2017backlash}. Those subjected to dehumanization may often experience negative emotions, develop more strained inter-group relationships, and may respond with reciprocal hostility \cite{kteily2017backlash}. To mitigate such dehumanization, prior work has suggested promoting inter-group contact, challenging hierarchical views of humanity, and emphasizing shared identities. Additionally, highlighting the similarities between groups can reduce subtle forms of dehumanization \cite{kteily2017backlash}. In fact, prior research has found employing these strategies in the virtual world can be effective in combating biased behaviors in the real world ~\cite{mulak2021virtual,peck2013putting, breves2020reducing, sahab2024contact, mckeown2017contact}. Thus, the incorporation of neurodiverse personas and behaviors into interactive AI agents can be a critical next step in combating dehumanization through the normalization of diverse communication styles.  

\subsection{Worlds Collide: How Virtual Interactions Impact Reality}
Contact hypothesis theorizes that people’s prejudices toward particular social groups may be reduced through contact with the group, and a systematic review found it does typically reduce prejudice ~\cite{paluck2019contact}, though there are notable exceptions and limitations (such as self-segregation) which may impact its effectiveness in the real world~\cite{mckeown2017contact}. However, prior work suggests contact with autistic people and knowledge of autism may improve autism acceptance among others. One study found familiarity with an autistic individual may decrease non-autistic people’s negative perceptions of autism ~\cite{dickter2021effects}, and another found providing autism acceptance training to non-autistic people reduces their explicit biases, improves their understanding of autism, and increases their interest in engaging with autistic people~\cite{jones2021effects}.

Researchers have examined the ways in which contact hypothesis occurring in digital spaces may impact people’s attitudes in real life, and have found diversity in video games may lead to more accepting attitudes amongst gamers towards diverse groups ~\cite{mulak2021virtual}. Similarly, other prior works have found that using racially diverse avatars led to less racially biased behavior in real world interactions ~\cite{peck2013putting}, and even non-playable characters of diverse backgrounds may decrease users’ explicit biases ~\cite{breves2020reducing}. In another study, researchers uncovered that conversational agents facilitating contact can improve inter-group attitudes even among groups that have a long history of conflict ~\cite{sahab2024contact}. This also shows the important role conversational agents can play in addressing the common shortcomings of the contact hypothesis, which is strongly dependent on the nature of the contact as negative contact increases prejudice ~\cite{mckeown2017contact}.

Even though there are benefits to promoting more diverse interactions online, our study evidences that neurotypical communication standards remain dominant in human-like robots and chatbots. Consequently, neurotypical people rarely gain opportunities to see autistic traits and characteristics as inherently human, rather than ``robotic.” Such exclusion reinforces the harmful perception that certain behaviors belong in the realm of machines, as they are not represented in systems designed to be human-like, compounding the marginalization of autistic individuals and other minority groups. For example, labeling particular behaviors as distinctly “human” versus “AI-like” can lead to the marginalization of individuals perceived to be using AI \cite{hohenstein2023artificial}. These issues can have dire consequences, such as job loss or false accusations of plagiarism, which disproportionately affect marginalized communities \cite{giray2024problem}. Thus, the absence of neurodiverse representations of humanness in such technologies is particularly concerning as they promote the neuronormative belief that there is only one right way to be a human~\cite{benson2023perplexing}. As media representations play a major role in autism acceptance in society ~\cite{mittmann2024portrayal}, it is important to ensure the technologies we create do not perpetuate biases regarding social norms that often dehumanize autistic people.

\subsection{Beyond Conference Policies: Other Recommendations for Systemic Changes}
Despite the recent push in publication standards at AI conferences to prioritize ethical considerations through means such as using ethics statements in submissions, we note that the broader ethical concerns of their work were not discussed by our participants. Even after exploring the conceptions of humanness implemented in their technologies, when asked about the ways these conceptions may influence the perceptions of human interactions in the real world or with AI held by others, the majority of participants (n=12) believed their work would have no impact on the way researchers building upon their work and their end-users view communication behaviors among humans. Similarly, the model biases and their impact were mentioned more frequently by our HCI-focused participants compared to our AI-focused ones, with the latter focusing solely on standard metrics such as precision to gauge the effectiveness of systems. This shows that despite the efforts to encourage more ethical work, AI researchers and engineers still struggle with understanding or explaining the broader ethical considerations that may arise from their work.

\subsubsection{Not an Afterthought: Centering Ethics in AI Education}
While AI technologies impact the lives of many humans either directly or indirectly, the creators of such systems have a tendency to view human concerns as beyond the scope of their responsibilities. In our study, participants frequently expressed sentiments such as “we are [machine learning] people, not HCI people” (P11), and referring to ethical concerns as the responsibility of others such as the product team (P1, P4). Such beliefs are mirrored even in our education system, with many AI educators having conflicting and contradictory thoughts toward ethics ~\cite{kamali2024ai}, and many universities separating those who work on “ethical” or “human-centered AI” from those who work on more “technical” projects through the creation of distinct departments or programs. This separation of “technical”  and “human-centered” perspectives may result in the former group receiving inadequate training in identifying ethical issues. For example, prior work has found AI courses hosted on YouTube neglected discussing ethics in favor of more technical content ~\cite{engelmann2024visions}, and undergraduate computer science students believe ethics are not prioritized, valued, or reward in their training or job prospects ~\cite{darling2024not}. 

In order to promote more ethical work, it is important for us to critically examine the separation of what we consider to be “ethical” or “responsible” AI from other forms of AI. Other engineering professions have standardized professional codes, legally binding standards, and license examinations, that prioritize safety and ethical concerns for all engineers. There is no such thing as an “ethical” structural engineer, for example, so why do we have that separation in AI? If we shrugged off ethical shortcomings as easily in other professions, we would need to exercise caution in only walking in buildings made by “ethical” engineers, for instance, to avoid having them spontaneously collapse under our feet. Yet, in the technical realm, the safety, well-being, and other ethical concerns of the broader general public are routinely sacrificed to be the “first” or “best” at releasing a product or a feature without any ethical examinations, despite the magnitude and scale of their impact on others’ lives sometimes being far larger than that of a single building. Perhaps, it is time we start a more thorough integration of ethics in our training of future makers, our evaluation of current makers and leaders, and the standards we all must uphold. 

\subsubsection{Organizational Support and Priorities}
While our participants expressed an interest in making their technologies more accessible, particularly for users who are international or have physical disabilities, they often mentioned being limited by their organization's support and priorities. For example, P3 noted accessibility was not a strong suit for their company, and reported having to ``push" to implement their considerations to enhance user experiences. They described having limited control over the overall design of their products, even mentioning that they had more flexibility in their university projects. In contrast, P15 responded having more support such as established standards for accessibility at their organization, and on-going projects that focused specifically on accessibility. These experiences show the impact organizational priorities can have on promoting or hindering more inclusive design, as many participants reported technical features were a bigger priority than user experiences for their organizations. Yet, there is a notable lack of clearly defined standards for these organizations and, consequently, any accountability for failing to adhere to them. While researchers have attempted to standardize guidelines for improving the accessibility of systems such as robots and chat-bots, as noted by prior work and reiterated by our own findings, these are not well-known even among the people who create these technologies ~\cite{qbilat2021proposal,stanley2022chatbot}.

\subsubsection{Fostering Diverse Teams for Socially Aware Perspectives}
An example of an ethical rule from the National Society of Professional Engineers (NSPE) Code of Ethics states that engineers shall “perform services only in areas of their competence” ~\cite{van2023ethics}. Applying such a rule in computing would require us to pursue interdisciplinary collaborations more fervently, as our work directly impacts the lives and livelihoods of communities we may not be familiar with but who may be the area of expertise of other trained professionals.

Indeed, we found the ethical and accessibility considerations that were mentioned by our participants were often related to their own knowledge and familiarity with the groups they believed would be impacted by their work. For example, P1 discussed accessibility considerations for blind/low vision users due to having a personal connection with a blind family member. Additionally, P2 mentioned implementing trauma-informed features in their work due to their experiences working with a minority community focusing on tasks that were often traumatizing. This highlights the importance of our other recommendation, which is fostering diversity in teams and, in support of the findings of prior work, encouraging greater community involvement in research studies ~\cite{birhane2019algorithmic, rizvi2024robots, begel2020lessons}. We must give communities decision-making power while developing technologies that impact them, so they are not treated as token minorities and are able to shape our work ~\cite{rizvi2024robots}. This means ensuring they have the power to change the design, implementation, and outcome of our work instead of merely including their perspectives towards the end of the developmental cycle.

\subsubsection{Changing How We Define Success}
In our study, many participants linked the humanization of their systems to increased engagement and better user experiences overall, echoing the sentiments of prior work ~\cite{chandra2022or, van2020human}. However, the humanization of technological agents raises ethical concerns toward the trustworthiness of such systems. For example, humanized chatbots can blur the boundary between the real and the virtual worlds, prompting people to trust misinformation \cite{maeda2024human} or even to perceive these bots as genuinely human. This distortion has already culminated in tragedy, as in the case of a teenager who developed a fraught, “inappropriate” relationship with a chatbot, became increasingly isolated from his family, and ultimately took his own life \footnote{\url{https://www.cnn.com/2024/10/30/tech/teen-suicide-character-ai-lawsuit/index.html}}. Furthermore, this humanization may be unnecessary for certain groups of users such as autistic people, who may prefer interacting with simpler systems ~\cite{robins2006does}. This leaves us with an important ethical consideration- do the benefits of humanizing technologies for certain users outweigh the harms this approach may cause to others? Perhaps we need to consider success metrics that go beyond user satisfaction, and think more deeply about the impact of our work on the lives of people we may inadvertently be marginalizing from solely focusing on the needs of the majority.

\section{Conclusion}
While there is a growing interest in humanizing AI agents such as robots and chatbots, our findings reveal this humanization is often done at the expense of autistic people’s preferences and broader inclusion in society. The creators of human-like AI who participated in our study displayed a clear preference for implementing neurotypical standards of communication in their work and often did not consider how this interpretation of humanness may impact their end-users perceptions of neurodivergent individuals. This is especially concerning as increasing diversity in the virtual world, and more positive representations of autism in the media have helped reduce explicit biases ~\cite{peck2013putting, jones2021effects, mittmann2024portrayal} among participants in previous studies, suggesting the impact more positive representations of diverse communication styles may have on autism inclusion in society. However, traits and behaviors commonly preferred by autistic people were often compared to non-human entities such as robots, illustrating the ways in which communication norms explicitly dehumanize autistic people. We encourage a deeper inclusion of community perspectives, a more thorough integration of ethics, clearly defined standards, and accountability for organizations in upholding these standards to mitigate similar biases in future work.

\bibliographystyle{ACM-Reference-Format}
\bibliography{custom}
\appendix
\section{Appendix}
\label{Appx}
\textcolor{red}{Note: in attempts to avoid response bias, we have included extra questions in our surveys and interviews. These are denoted by `*' if included in the Appendix. }
\subsection{Surveys}
\subsubsection{Survey 1: Demographic Questions}
\label{Survey1}

\begin{enumerate}

\item \textbf{What is your job official title?}\\[6pt]
\noindent\rule{0.9\linewidth}{0.4pt}  

\item \textbf{What industry do you work in?}
\begin{itemize}
  \item Academia
  \item Industry
  \item Other: \rule{0.5\linewidth}{0.4pt}
\end{itemize}

\item \textbf{Please specify the gender with which you most closely identify.}
\begin{itemize}
  \item Woman
  \item Man
  \item Non-binary
  \item Prefer not to answer
  \item Prefer to self-describe: \rule{0.5\linewidth}{0.4pt}
\end{itemize}

\item \textbf{What is your age?}
\begin{itemize}
  \item 18--22
  \item 23--30
  \item 31--40
  \item 41--50
  \item 51--60
  \item Over 60
  \item Prefer not to answer
\end{itemize}

\item \textbf{Please specify your race/ethnicity.}
\begin{itemize}
  \item White
  \item Hispanic or Latino
  \item Black or African American
  \item Native American or American Indian
  \item South Asian
  \item Southwest Asian or North African
  \item East Asian
  \item Southeast Asian or Pacific Islander
  \item Prefer to self-describe: \rule{0.5\linewidth}{0.4pt}
  \item Prefer not to answer
\end{itemize}

\item \textbf{Parental Status *}
\begin{itemize}
  \item I have or had children that I currently support and/or raise by myself, 
        with a partner, or as part of a team or family
  \item I do not have children that I currently support and/or raise by myself, 
        with a partner, or as part of a team or family
  \item Prefer to self-describe: \rule{0.5\linewidth}{0.4pt}
  \item Prefer not to answer
\end{itemize}

\item \textbf{Marital status *}
\begin{itemize}
  \item Single
  \item Common Law Marriage
  \item Married
  \item Separated
  \item Divorced
  \item Widowed
  \item Legal
  \item Prefer to self-describe: \rule{0.5\linewidth}{0.4pt}
  \item Prefer not to answer
\end{itemize}

\item \textbf{6-Question Disability Questionnaire}

\begin{enumerate}[label*=\arabic*.]
  \item Are you deaf, or do you have serious difficulty hearing?
  \item Are you blind, visually impaired, or do you have serious difficulty seeing, even when wearing glasses?
  \item Because of a physical, mental, or emotional condition, do you have serious difficulty 
        concentrating, remembering, or making decisions?
  \item Do you have serious difficulty walking or climbing stairs?
  \item Do you have difficulty dressing or bathing?
  \item Because of a physical, mental, or emotional condition, do you have difficulty doing 
        errands alone such as visiting a doctor’s office or shopping? (15 years old or older)
\end{enumerate}

\item \textbf{Sexual Orientation *}
\begin{itemize}
  \item Asexual
  \item Bisexual
  \item Pansexual
  \item Gay
  \item Heterosexual
  \item Lesbian
  \item Queer
  \item Prefer to self-describe: \rule{0.5\linewidth}{0.4pt}
  \item Prefer not to answer
\end{itemize}

\item \textbf{What is the highest degree or level of school you have completed?}
\begin{itemize}
  \item Some high school credit, no diploma or equivalent
  \item Less than high school degree
  \item High school graduate (high school diploma or equivalent including GED)
  \item Some college but no degree
  \item Associate's degree
  \item Bachelor's degree
  \item Advanced degree (e.g., Master's, doctorate)
  \item Prefer not to answer
\end{itemize}

\item \textbf{Which one of the following includes your total HOUSEHOLD income for last year, before taxes?}
\begin{itemize}
  \item Less than \$10,000
  \item \$10,000 to under \$20,000
  \item \$20,000 to under \$30,000
  \item \$30,000 to under \$40,000
  \item \$40,000 to under \$50,000
  \item \$50,000 to under \$65,000
  \item \$65,000 to under \$80,000
  \item \$80,000 to under \$100,000
  \item \$100,000 to under \$125,000
  \item \$125,000 to under \$150,000
  \item \$150,000 to under \$200,000
  \item \$200,000 or more
  \item Prefer not to answer
\end{itemize}

\end{enumerate}
\subsubsection{Survey 2}
\label{Survey2}
\begin{enumerate}
  \item \textbf{Imagine your team switches to a remote-optional work environment and 
  now allows your team members to have their videos off during meetings. 
  How do you feel about this new policy?}
  \begin{itemize}
    \item Strongly Disagree
    \item Somewhat Disagree
    \item Neither agree nor disagree
    \item Somewhat Agree
    \item Strongly Agree
  \end{itemize}

  \item \textbf{Imagine your team does not allow people to have conversations in the open 
  work spaces, and requires them to use designated spaces. How do you feel about this 
  change in work policies?}
  \begin{itemize}
    \item Strongly Disagree
    \item Somewhat Disagree
    \item Neither agree nor disagree
    \item Somewhat Agree
    \item Strongly Agree
  \end{itemize}

\end{enumerate}
\subsection{Interviews}
\subsubsection{Interview 1}
\label{Int1}
\begin{enumerate}[label=\arabic*.]
    \item Tell me about a recent project you worked on where you built an AI system that others would potentially use.
    \begin{enumerate}[label=\alph*.]
        \item When did you know you’ve reached the end of the design process for your system?
        \begin{enumerate}[label=\roman*.]
            \item When the current assignment is complete.
        \end{enumerate}
        \item When did you know you’ve reached the end of the design process for your Robot/chatbot/AI agent?
        \item What was the overall purpose or objective for your system?
        \item How did you decide the physical characteristics (including UI) of the bot?
        \begin{enumerate}[label=\roman*.]
            \item What do you think would happen if you chose something else?
        \end{enumerate}
        \item How did you decide the behavioral and conversational characteristics of the bot?
        \begin{enumerate}[label=\roman*.]
            \item What do you think would happen if you chose something else?
        \end{enumerate}
    \end{enumerate}

    \item What are the metrics you use for evaluating the performance of your system?
    \item How did you know that your work was effective? (ask if they created a solution to a problem, otherwise skip).
    \begin{enumerate}[label=\alph*.]
        \item Can you tell me about a time when a user interacted with your system and got frustrated or upset?
        \item What are some things that may make a user frustrated or not want to interact with your system?
    \end{enumerate}

    \item Can you walk me through how or why you chose this specific mode of user interaction to reach your objective or solve your problem?

    \item What real-world interactions can you think of that resemble your user interaction?
    \begin{enumerate}[label=\alph*.]
        \item If they say no, ASK - Can you think of any other interaction that might be similar to the interaction you developed?
        \item What are the identities of the people in these interactions?
        \item Explain the interaction for me.
        \begin{enumerate}[label=\roman*.]
            \item What would it look like if one of the people was autistic/ADHD/dyslexic?
            \item How would the interaction change?
        \end{enumerate}
    \end{enumerate}

    \item What behavioral or communicational changes could you make to the bot that would impact the user’s interaction with the bot?
    \begin{enumerate}[label=\alph*.]
        \item How would that be different with a person who is funny?
        \item How would that impact the user’s interactions with folks who are and aren’t funny/don’t use humor in interactions?
    \end{enumerate}

    \item What conclusions would researchers who are building off of your work reach or assume about human interactions, behavior, or AI? What about end-users? What conclusions might they reach?
    \begin{enumerate}[label=\alph*.]
        \item What are the best practices in bot design you have learned in your work?
        \item How do end-users’ interactions differ from those with other bots?
    \end{enumerate}

    \item In your opinion, should we design chatbots, robots, or other AI agents to meet the needs of various community groups?
    \begin{enumerate}[label=\alph*.]
        \item Who are the people most likely to use your chatbot?
        \item What would happen if someone from [specific group] used it?*
        \item What would happen if you adjusted the bot to work with multiple user groups? Or specific user groups?*
        \item What kind of accessibility features did you consider in your chatbot?*
        \item Should we design for people whose mental and neurological abilities may be considered atypical such as folks with autism, ADHD, and dyslexia?
        \begin{enumerate}[label=\roman*.]
            \item Limitations?
        \end{enumerate}
        \item Should we design for different age groups?*
    \end{enumerate}

    \item Is there anything else you’d like to say or add in regards to your experience building AI entities?
\end{enumerate}

\subsubsection{Interview 2}
\label{Int2}
\begin{enumerate}[label=\arabic*.]
    \item How would you describe your current professional position?
    \item In your opinion, what experiences from your life greatly impacted or led you to your interest in your current career?
    \item What made you stay or continue to pursue your current career?
    \item Tell me about a time when your personal background overlapped with your current position or your work.
\end{enumerate}

\textbf{Intersectionality Questions}
\begin{enumerate}[label=\arabic*.]
    \item Has an ism (e.g., racism, sexism, ableism, homophobia) ever impacted you during your computer science journey?*
    \begin{enumerate}[label=\alph*.]
        \item Has an ism ever impacted you in your life?
        \item If no, ask: At work, do you interact with a number of people who have a similar racial, ethnic, or gender background as you on a regular basis?
    \end{enumerate}
    \item In your opinion, what is intersectionality?*
    \begin{enumerate}[label=\alph*.]
        \item Has it ever impacted your work or research? If so, how?
    \end{enumerate}
    \item Tell me about a time when you used or experienced intersectionality in your work/research.*
    \begin{enumerate}[label=\alph*.]
        \item If non-tech, let them provide it. If research-related, ask: Do you have any examples related to your current position?
    \end{enumerate}
    \item Tell me about a time when you used or experienced a push for inclusivity in your work/research.*
    \item Tell me a time when intersectionality has impacted you at work.*
    \item Have you had any training in working with or building tools that support inclusivity?
    \begin{enumerate}[label=\alph*.]
        \item If so, what did the training look like? Can you describe it?
        \item Did it help you?
    \end{enumerate}
    \item Have you had any support in working with or building tools that support inclusivity?*
    \begin{enumerate}[label=\alph*.]
        \item If so, what did the support look like? Can you describe it?
        \item Did it help you?
    \end{enumerate}
\end{enumerate}

\textbf{Situational Awareness}
\begin{enumerate}[label=\arabic*.]
    \item Suppose your new colleague shows up wearing headphones to a research conference and continues wearing them even while presenting their work. How would you approach a conversation with them?
    \begin{enumerate}[label=\alph*.]
        \item What if your coworker mentions they’re wearing it because the room is too loud?
    \end{enumerate}
    \item Imagine that your company celebrates its anniversary during Ramadan. Ramadan is a Muslim holiday where Muslims refrain from eating and drinking from sunrise to sunset. This year is their 50th anniversary, and the plan is to have a week full of free buffet-style lunch and brunch banquets over the weekend. However, about three coworkers have expressed concern that they won’t be able to fully participate due to the way things are planned to happen.
    \begin{enumerate}[label=\alph*.]
        \item What is the problem from the coworkers' viewpoint?*
        \item How might this problem be solved?*
    \end{enumerate}
    \item Suppose your research assistant doodles during meetings, and your supervisor approaches you about the lack of professionalism exhibited by your RA. How would you address this situation?
    \item Max and Alex are working on a research project and are expected to provide feedback to each other to ensure the project stays on track. [Share a conversation transcript.]
    \begin{enumerate}[label=\alph*.]
        \item Why is Alex responding this way?
        \item Why is Max responding this way?
        \item How might you respond as their supervisor?
    \end{enumerate}
\end{enumerate}
\end{document}